# Crystallographically oriented magnetic $ZnFe_2O_4$ nanoparticles synthesized by Fe implantation into ZnO

Shengqiang Zhou, K. Potzger, H. Reuther, G. Talut, F. Eichhorn, J. von Borany, W. Skorupa, M. Helm and J. Fassbender

*Institute of Ion Beam Physics and Materials Research, Forschungszentrum Rossendorf, P.O. Box 510119, 01314 Dresden, Germany*

Abstract

In this paper, a correlation between structural and magnetic properties of Fe implanted ZnO is presented. High fluence $Fe^+$ implantation into ZnO leads to the formation of superparamagnetic α-Fe nanoparticles. High vacuum annealing at 823 K results in the growth of α-Fe particles, but the annealing at 1073 K oxidized the majority of the Fe nanoparticles. After a long term annealing at 1073 K, crystallographically oriented $ZnFe_2O_4$ nanoparticles were formed inside ZnO with the orientation relationship of $ZnFe_2O_4(111)[110]//ZnO(0001)[11\underline{2}0]$. These $ZnFe_2O_4$ nanoparticles show a hysteretic behavior upon magnetization reversal at 5 K.

## I. Introduction

Diluted magnetic semiconductors (DMS) have recently attracted huge research attention because of their potential application for spintronics devices [1, 2]. In DMS materials, transition or rare earth metal ions are substituted onto host cation sites and are coupled with free carriers to yield ferromagnetism via indirect interaction [3, 4]. ZnO doped with Fe was found to be ferromagnetic at room temperature [5-8]. However, the origin of the observed ferromagnetism in transition metal doped ZnO is still controversial, e.g. ferromagnetic clusters [9-12], or extrinsic reasons [13, 14] are discussed. Moreover superparamagnetic Co and Ni nanoclusters are formed by ion implantation into ZnO [15, 16] and $TiO_2$ [17, 18], respectively. To clarify the origin of the observed ferromagnetism, a careful correlation between structure and magnetism should be established by sophisticated characterization methods. Synchrotron radiation based x-ray diffraction (SR-XRD) is a powerful tool to detect small precipitates, while element selective measurement of the magnetic properties is desired, e.g. using conversion electron Mössbauer spectroscopy (CEMS). In this paper, employing the above-mentioned methods together with superconducting quantum interference device (SQUID) magnetometry, we report investigations on the evolution of structural and magnetic properties of Fe implanted ZnO upon high vacuum annealing. Moreover the possibility to form crystallographically oriented Zn-ferrite embedded in ZnO is demonstrated. With respect to the crystal symmetry and lattice mismatch, our results suggest that other ferrites, which have been epitaxially grown onto MgO, $SrTiO_3$, and $Y_{0.15}Zr_{0.85}O_2$, and exhibit rich magnetic properties [19], could also be embedded inside single crystalline ZnO.

## II. Experiments

Hydrothermally grown, commercially available ZnO single crystals were implanted with a $^{57}Fe$ fluence of $4\times10^{16}$ cm$^{-2}$ at 623 K. The implantation energy of 180 keV yielded a projected range of $R_P$=83±35 nm (TRIM code). The post-implantation thermal annealing was performed in a high vacuum ($p<10^{-7}$ mbar) furnace from 823 K to 1073 K for either 15 minutes or 3.5 hours. Virgin and implanted samples were investigated using Rutherford backscattering/channeling spectrometry (RBS/C), SR-XRD with an x-ray wavelength of 0.154 nm at Rossendorf beamline (ROBL) at the ESRF, SQUID (Quantum Design MPMS) magnetometry, room temperature CEMS and scanning electron microscopy (SEM).

III. Results and discussion

## A. Secondary phase evolution

Fig. 1a shows the SR-XRD patterns for the as-implanted and annealed samples. For the as-implanted sample, crystalline α-Fe nanoparticles were observed, and no other crystalline Fe-oxide ($Fe_2O_3$, $Fe_3O_4$, and $ZnFe_2O_4$) particles were detected. The crystallite size is calculated using the Scherrer formula [20]. The peak amplitude and crystallite size are compared in table I. After 823 K and 15 min annealing, larger and more Fe nanoparticles are formed reflected by an increase and the sharpening of the corresponding peak at 44.4° in the 2θ-θ scan (Fig. 1a). After 1073 K and 15 min annealing, the Fe(110) peak almost disappeared and the sample already shows an indication for the presence of crystalline $ZnFe_2O_4$. After 3.5 hours annealing at 1073 K, crystalline and oriented $ZnFe_2O_4$ particles are clearly identified. The inset shows a zoom of the Fe(110) peak to show the development of Fe nanoparticles more clearly.

## B. Magnetism evolution

Fig. 1b shows the magnetization versus field reversal (M-H) at 5 K. Magnetic hysteretic loops are observed for the as-implanted sample, which contains α-Fe nanoparticles. After annealing at 823 K for 15 min, the ferromagnetism is enhanced, i.e. Fe nanoparticles are growing in size and amount. However, after annealing at 1073 K for 15 min, no hysteresis loop is observed.

Probably the majority of Fe particles were oxidized to some amorphous nonmagnetic compound. The magnetism evolution is in a good agreement with the XRD measurement. After annealing at 1073 K for 3.5 hours, the hysteretic behavior is observed again, which – according to the SR-XRD - cannot originate from metallic Fe but from $ZnFe_2O_4$ nanoparticles. Bulk $ZnFe_2O_4$ exhibiting normal spinel structure is a weak antiferromagnet with a Neél-temperature of 10.5 K [21], while epitaxial $ZnFe_2O_4$ thin films exhibit a higher Néel temperature of 43 K [22]. However recent experiments reveal that nanocrystalline $ZnFe_2O_4$ shows ferrimagnetic behaviour [23-31]. The explanation for such behaviour is the partial inversion of the spinel structure, i.e. the additional occupation of tetrahedral A sites by Fe and octahedral B sites by Zn leading to a strong superexchange coupling of the intra-sublattice Fe ions. Therefore, the $ZnFe_2O_4$ nanoparticles (see Fig. 1a) are responsible for the hysteretic behaviour upon magnetization reversal observed for the long term annealed sample.

In order to confirm the superparamagnetic properties of the embedded nanoparticles, zero-field cooled (ZFC) and field cooled (FC) magnetization vs. temperature measurements were performed using SQUID. Figure 1c shows the ZFC/FC magnetization curves in a 5 mT field. For all samples, although there is a difference in magnetic moment, ZFC curves show a gradual increase (deblocking) at low temperatures, and reach a plateau at a particular temperature of $T_{max}$, while all FC curves continue to increase with decreasing temperature. The shape of the ZFC/FC curves are general characteristics of magnetic nanoparticle systems [32], i.e. magnetic nanoparticles are the origin of the ferromagnetism in all the samples. The increase of $T_{max}$ after annealing at 823 K for 15 min confirms the growing of α-Fe nanoparticles. After annealing at 1073 K for 3.5 h, $ZnFe_2O_4$ crystallites with an average grain size of 20 nm were formed. However the degree of inversion (transition from $Fe^{3+}$ to $Fe^{2+}$) is decreased with increasing crystallites [30], therefore comparing with the grain size of 6.6 nm and 14.8 nm in Ref. 30, the rather bigger grain size of $ZnFe_2O_4$ crystallites (20 nm) results in a small magnetic moment and a low $T_{max.}$ Also the effective anisotropy constant of $ZnFe_2O_4$ is lower than that of α-Fe [29], which could explain the lower $T_{max}$ than that of Fe nanoparticles [32]. The magnetic and structural properties were compared in Table I.

One way to link the electronic with the magnetic properties of Fe is given by CEMS. The results obtained do fully support the findings by XRD and SQUID-magnetometry. As presented in Fig. 2 and Table II, the as implanted sample shows dominant $Fe^{3+}$ and $Fe^{2+}$ states that are associated with disperse Fe ions. A small fraction of 12.5 % of the implanted $^{57}Fe$ develops a magnetic sextet associated with α-Fe nanoparticles according to its isomer shift of 0 mm/s with respect to the reference sample. Upon annealing at 823 K, the intensity of the sextet increases up to 18.2 % while the fraction of $Fe^{2+}$ decreases, suggesting the growth of the α-Fe nanoparticles and the recovery of lattice defects. Moreover, the value for the magnetic hyperfine field $B_{HF}$ increases upon annealing and moves closer to the known value for bulk α-Fe (33 T). The sextet disappears upon annealing for 15 min at 1073 K. Additionally a new doublet occurs around 0.37 mm/s exhibiting a large quadrupole splitting, while the non-split $Fe^{3+}$ line associated with disperse Fe is lowered. After 3.5 hours annealing, the CEMS pattern exhibits only one quadrupole split line typical for $ZnFe_2O_4$ at room temperature [24, 27-30]. The CEMS spectra of the latter two annealing steps systematically prove that the disappearance of the α-Fe peak in XRD is not related to dissolving but to the formation of $ZnFe_2O_4$.

## C. Crystallographically oriented $ZnFe_2O_4$

In Fig. 1a, the XRD pattern for the sample after 3.5 hours annealing at 1073 K shows only three peaks of $ZnFe_2O_4$ (222) (333) and (444), which means that the crystallites of $ZnFe_2O_4$ are highly oriented with respect to the host matrix. The surface orientation is $ZnFe_2O_4(111)//ZnO(0001)$. The crystallographical orientation of $ZnFe_2O_4$ was revealed by the XRD pole figure. Fig. 3 shows the pole figure for $ZnFe_2O_4(311)$. The radial coordinate is the angle ($\chi$) by which the surface is tilted out of the diffraction plane from 20° to 70°. The azimuthal coordinate ($\Phi$) is the angle of rotation about the surface normal. The pole figure shows poles of $ZnFe_2O_4(311)$ at $\chi$~29.5° and 58.5°, respectively, with sixfold symmetry. Since $ZnO(10\underline{1}1)$ ($2\theta$=36.25°) has a close Bragg angle with $ZnFe_2O_4(311)$ ($2\theta$=35.27°), the poles of $ZnO(10\underline{1}1)$ also show up at $\chi$~61.6° with much more intensities. The result is consistent with the theoretical $ZnFe_2O_4(311)$ pole figure viewed along [111] with rotation twins. The in-plane orientation relationship is $ZnFe_2O_4[110]//ZnO[11\underline{2}0]$. Moreover, a $2\theta$-$\theta$ scan was carried out for $ZnFe_2O_4(311)(622)$ and (220)(440), respectively, by tilting the sample with an angle of $\chi$ at an azimuthal position found by pole figure. The results are shown in Fig. 4, and confirm the crystallographical orientation of $ZnFe_2O_4$. The coherence length of crystallites is around 20 nm in the out-of-plane direction (Table I). The in-plane coherence length is evaluated to be also as large as 20 nm by measuring the diffraction of (311) at $\chi$~80°, nearly parallel with the surface [33]. Due to the fcc structure of $ZnFe_2O_4$ ($a$=0.844 nm), it is not difficult to understand its crystallographical orientation onto hcp-ZnO ($a$=0.325 nm) with twin-crystallites of $ZnFe_2O_4$ of an in-plane rotation by 60º (Fig.5). The lattice mismatch between $ZnFe_2O_4$ and ZnO is ~6%.

The orientation of $ZnFe_2O_4$ is also indirectly evidenced by RBS/channeling. The channeling spectra were collected by aligning the sample to make the impinging $He^+$ beam parallel with ZnO<0001> axis. $\chi_{min}$ is the channeling minimum yield in RBS/C, which is the ratio of the backscattering yield at channeling condition to that for a random beam incidence [34]. Therefore, the $\chi_{min}$ labels the lattice disordering degree, and an amorphous sample shows a $\chi_{min}$ of 100 %, while a perfect single crystal corresponds to a $\chi_{min}$ of 1-2 %. Fig. 6 shows the representative RBS/C spectra for different annealing procedures. The humps in the channeling spectra mainly come from the randomly located Fe and the lattice disordering due to implantation. $\chi_{min}$ is drastically decreased after annealing at 1073 K for 3.5 hours (Table I), which suggests the partial recovering of lattice damage, and the formation of an oriented secondary phase by thermal annealing [35]. In our case, the <111> axis of $ZnFe_2O_4$ is parallel to the ZnO<0001> axis, the direction of the impinging $He^+$ beam, which reduces the backscattering yield.

Additionally SEM was performed to check the sample morphology. Fig. 7a and 7b show the SEM results for the as-implanted and the annealed samples. After the annealing, the morphology was pronouncedly changed. Some speckle-like features with dimensions of 100 nm were formed on top of ZnO. Using energy dispersive x-ray (EDX), these features are Fe-riched islands, while in the flat area, no Fe is detectable (Fig. 7c). Together with above-mentioned SR-XRD and CEMS observations, these speckle-like Fe-riched features are most probably $ZnFe_2O_4$. This feature size is not necessary to be the same as the coherence length (crystallite size) as revealed by XRD since one feature can consist several crystallites, however SEM cannot distinguish them. Thus, Fe is partially diffused towards the sample surface during annealing, and therefore the $ZnFe_2O_4$ crystallites are located in the region near surface.

Generally, spinel ferrites ($MFe_2O_4$, M=Ni, Co, Fe, Mn, Zn) have a large variety of magnetic properties and have significant potential application in millimeter wave integrated circuitry and magnetic recording [19]. They all have an fcc structure with the lattice constant *a* of around 0.84 nm, and exhibit different magnetic properties depending on the chemical composition and cation site occupancy [36]. In the view of lattice mismatch, our results could suggest the epitaxy of spinel ferrites onto ZnO, and even a multi-layered $MFe_2O_4$/ZnO structure given the growth method compatibility by pulsed laser deposition or molecular beam epitaxy for both materials [19, 37]. The magneto-transport properties of inverted $ZnFe_2O_4$ nanoparticles are recently investigated. Ponpandian *et al.* discussed a hopping mechanism between $Fe^{3+}$ and $Fe^{2+}$ pairs present at the octahedral sites [38]. In Ref. 23, Shinagawa *et al.* reported ferromagnetic ZnO-Spinel iron oxide composites prepared by wet chemical process with a negative magnetoresistance of -0.35% at room temperature. Thus, a hybrid structure of spinel ferrites/semiconducting ZnO could be a potential candidate for spintronic devices. The crucial point about spin injection from the nanoparticles into the host material, however, is the quality of their interface. Although the high degree of crystalline orientation indirectly suggests a very sharp interface, local investigations have to be performed in order to clearly reveal the interface structure.

## IV. Conclusions

Magnetic and structural properties were correlated in Fe implanted ZnO crystals and we demonstrate the possibility to form crystallographically oriented Zn-ferrite with respect to the ZnO host single crystal. In the as-implanted samples, α-Fe nanoparticles are the origin of the ferromagnetism, 823 K annealing induced the growth of Fe particles, however 1023 K, 15 min annealing oxidized the main part of Fe particles. Zn-ferrite was formed after 3.5 hours annealing at 1073 K with the crystalline orientation of $ZnFe_2O_4(111)(110)//ZnO(0001)[11\underline{2}0]$. The Zn-ferrite nanoparticles are partially inverted, and a strong A-B superexchange interaction produces the ferrimagnetic coupling.

## Acknowledge

We thank Dr. N. Schell for technical support for SR-XRD measurements.

Fig captions

Fig. 1 (a) SR-XRD patterns (2θ-θ scans) of Fe implanted ZnO reveal the second phase development (from α-Fe to $ZnFe_2O_4$) upon annealing. (b) Magnetization versus field reversal revealing the magnetism evolution upon annealing; (c) ZFC/FC magnetization curves at 5 mT, which exhibit a typical characteristic of a magnetic nanoparticle system.

Fig. 2 CEMS after different annealing procedures. The arrows point the sextet from α-Fe.

Fig. 3. Pole figure of $ZnFe_2O_4$(511) reveals the crystallographical orientation of $ZnFe_2O_4$ and its twin crystallites.

Fig. 4. 2θ-θ scans of $ZnFe_2O_4$(311)(622) and (220)(440).

Fig. 5. Schematics for the crystallographical orientation of $ZnFe_2O_4$ onto ZnO.

Fig. 6. Representative RBS/channeling spectra with different annealing procedures.

Fig. 7. SEM shows the morphology of samples (a) rather flat surface in as-implanted sample, (b) Fe-riched features on the surface after 3.5 hours annealing at 1073 K. (c) EDX spectra at different spot of Fig. 5b (inset shows the higher energy parts of the spectra).

Table I Structure and magnetism evolution of Fe-implanted ZnO upon annealing

| Annealing | Peak area of Fe(110) and crystallite size | Peak area of $ZnFe_2O_4$(333) and crystallite size | Saturation moment at 5K | $T_{max}$ (ZFC at 5 mT) | $\chi_{min}$ (RBS/C) |
|---|---|---|---|---|---|
| As-impl. | 487 / 7.1 nm | - | 0.24 $\mu_B$/Fe | 146 K (Fe) | 63% |
| 823K, 15 m | 837 / 9.4 nm | - | 0.34 $\mu_B$/Fe | 200 K (Fe) | 57% |
| 1073K, 15m | - | - | - | - | 43% |
| 1073K, 3.5h | - | 462 / 20 nm | 0.14 $\mu_B$/Fe | 23 K ($ZnFe_2O_4$) | 29% |

Table II. Hyperfine parameters measured using CEMS for samples after different annealing treatment.

| Annealing Temperature and Time | $Fe^{3+}$ | | | $Fe^{2+}$ | | | $Fe^{0}$ | | |
|---|---|---|---|---|---|---|---|---|---|
| | FR[a] (%) | IS[b] (mm/s) | QS[c] (mm/s) | FR (%) | IS (mm/s) | QS (mm/s) | FR (%) | IS (mm/s) | $B_{HF}$[d] (T) |
| As impl. at 623 K | 32.8 | 0.53 | 0 | 23.2 | 0.96 | 0.58 | 12.5 | 0.06 | 30.5 |
| | | | | 31.5 | 0.78 | 1.29 | | | |
| 823 K / 15 min | 42.6 | 0.53 | 0 | 22.5 | 0.94 | 0.54 | 18.2 | 0.07 | 31.7 |
| | | | | 16.7 | 0.68 | 1.52 | | | |
| 1023 K / 15 min | 26.3 | 0.56 | 0 | - | - | - | 11.2 | 0.15 | 0 |
| | 62.4 | 0.37 | 0.95 | | | | | | |
| 1023 K / 3.5 h | 100 | 0.35 | 1.06 | - | - | - | - | - | - |

[a] Fraction

[b] Isomer shift

[c] Quadrupole splitting

[d] Magnetic hyperfine field

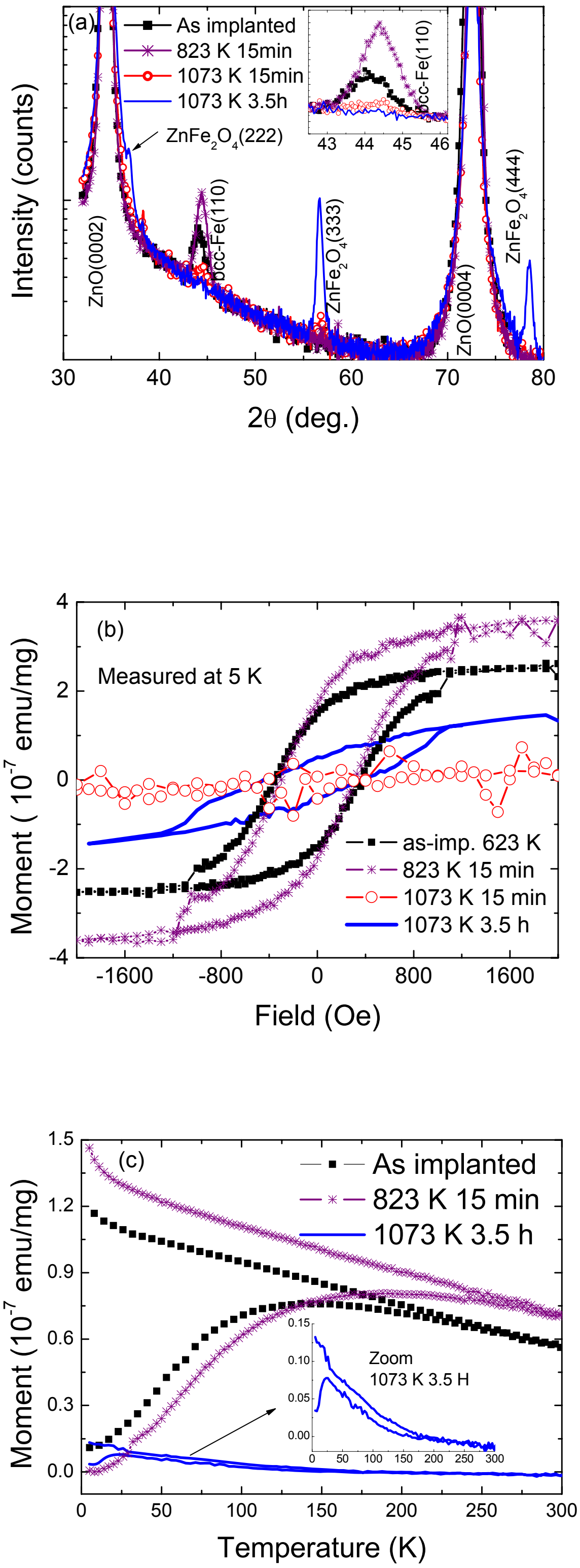

Fig. 1

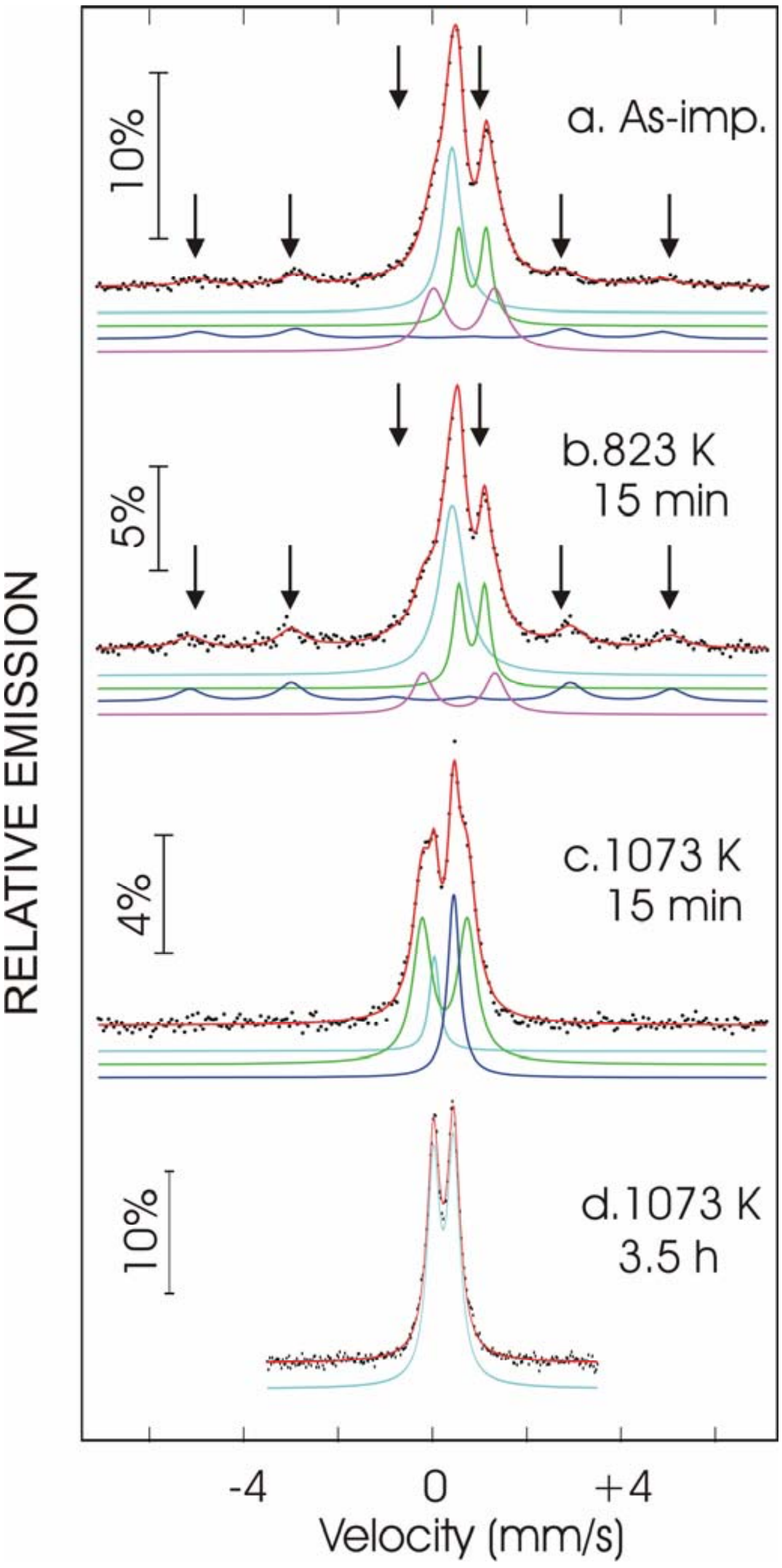

Fig. 2

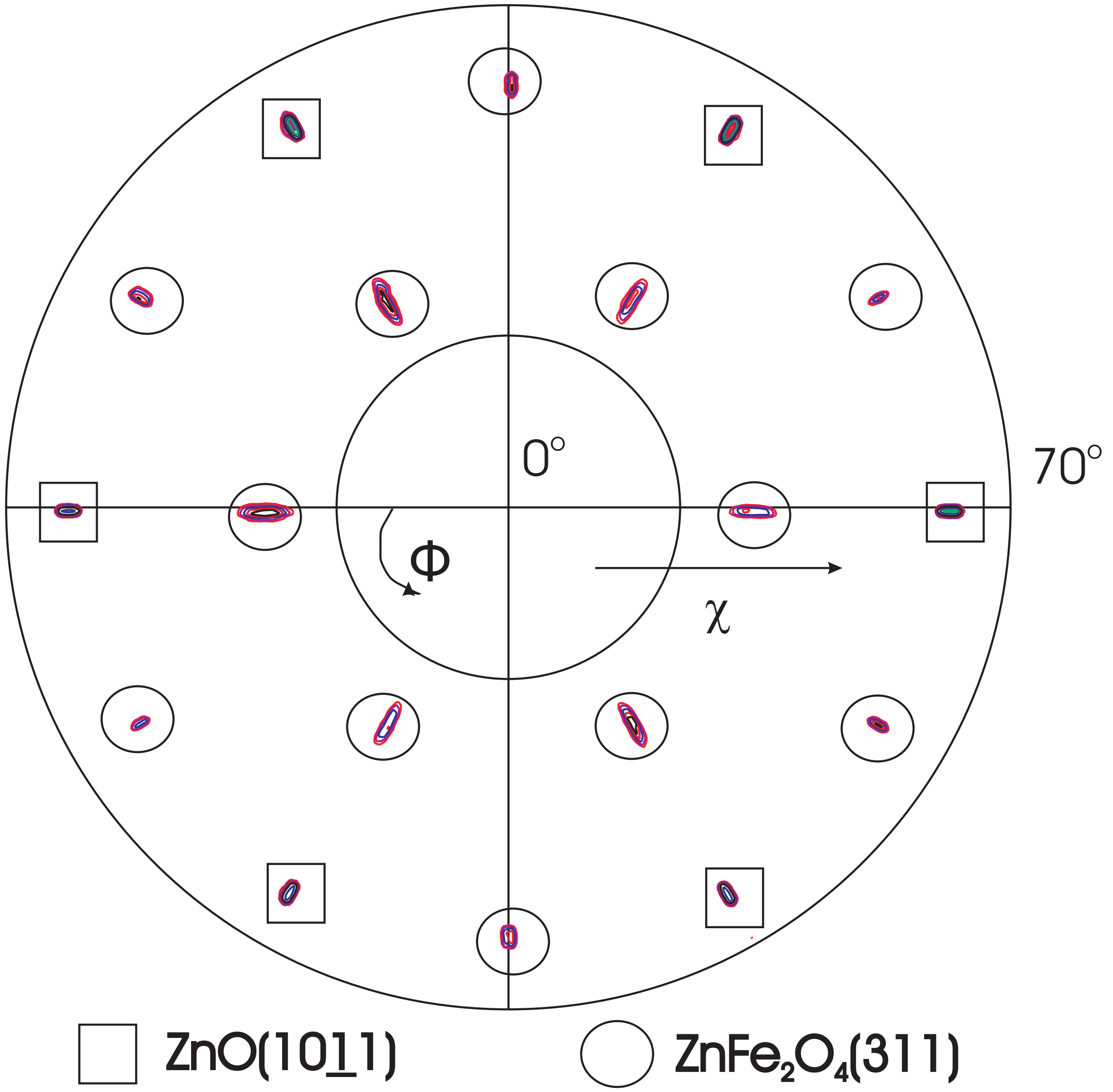

Fig. 3

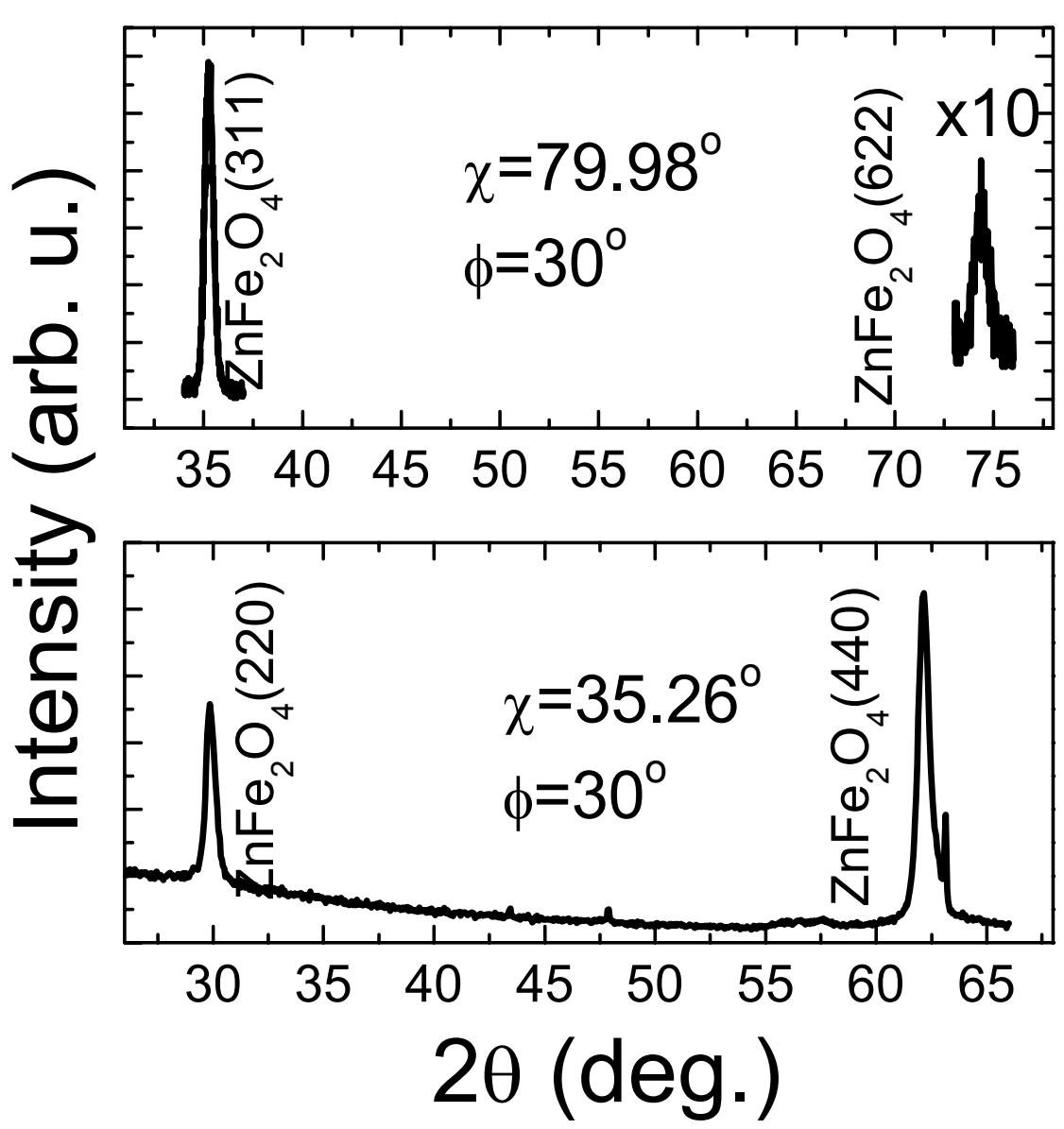

Intensity (arb. u.)
ZnFe2O4(311)
χ=79.98°
φ=30°
ZnFe2O4(622)
x10
35 40 45 50 55 60 65 70 75
ZnFe2O4(220)
χ=35.26°
φ=30°
ZnFe2O4(440)
30 35 40 45 50 55 60 65
2θ (deg.)

Fig. 4

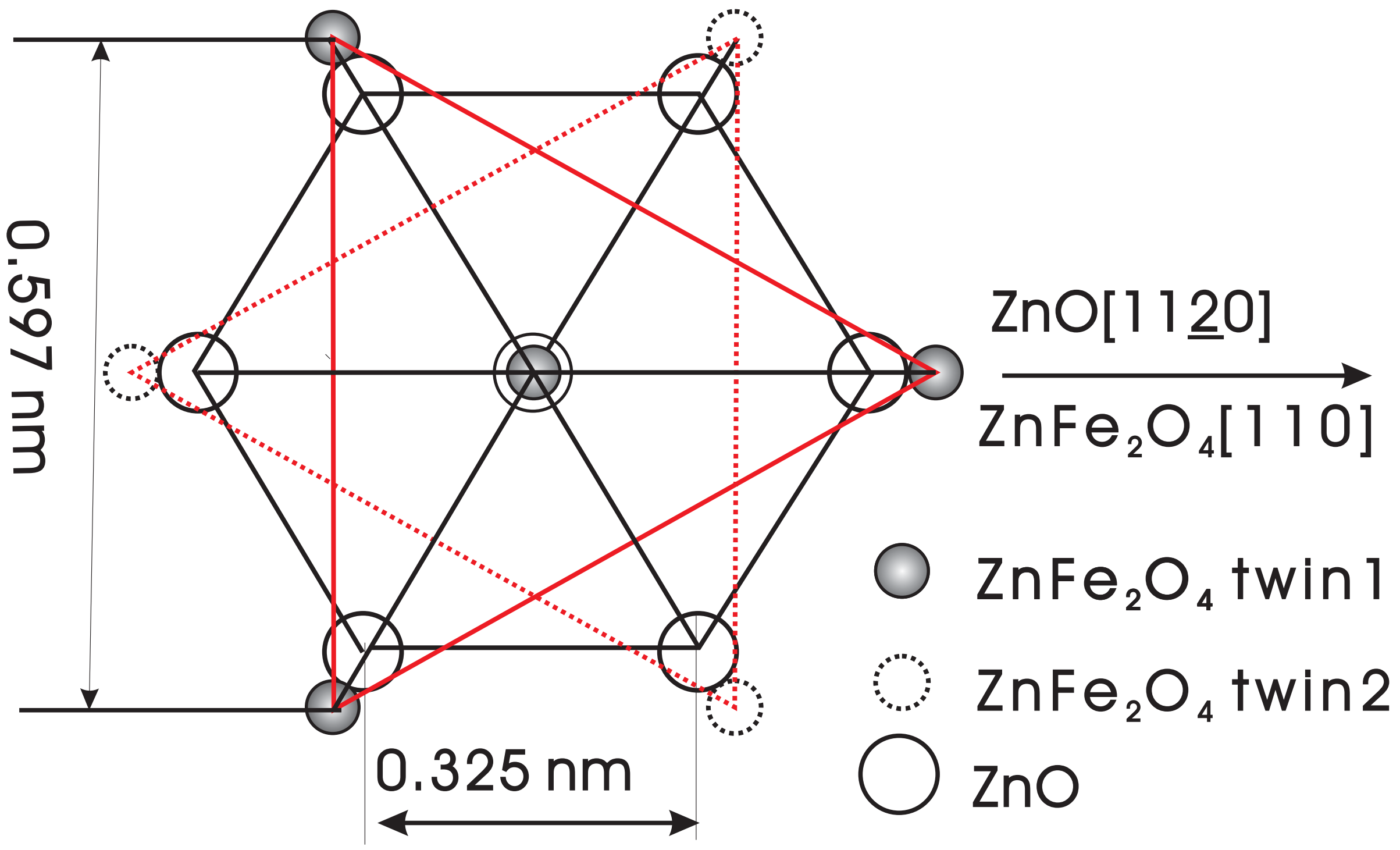

Fig. 5

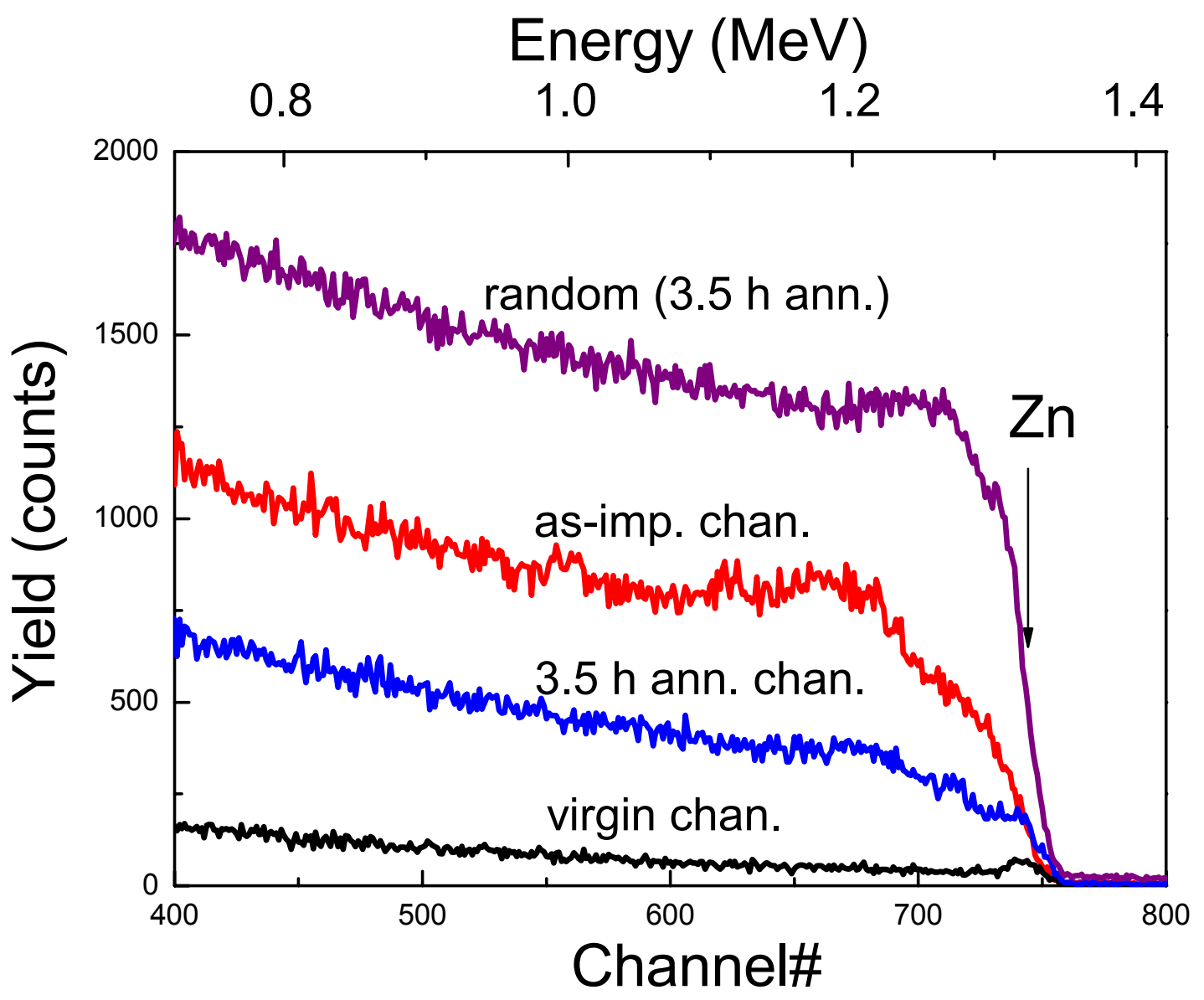

Fig. 6

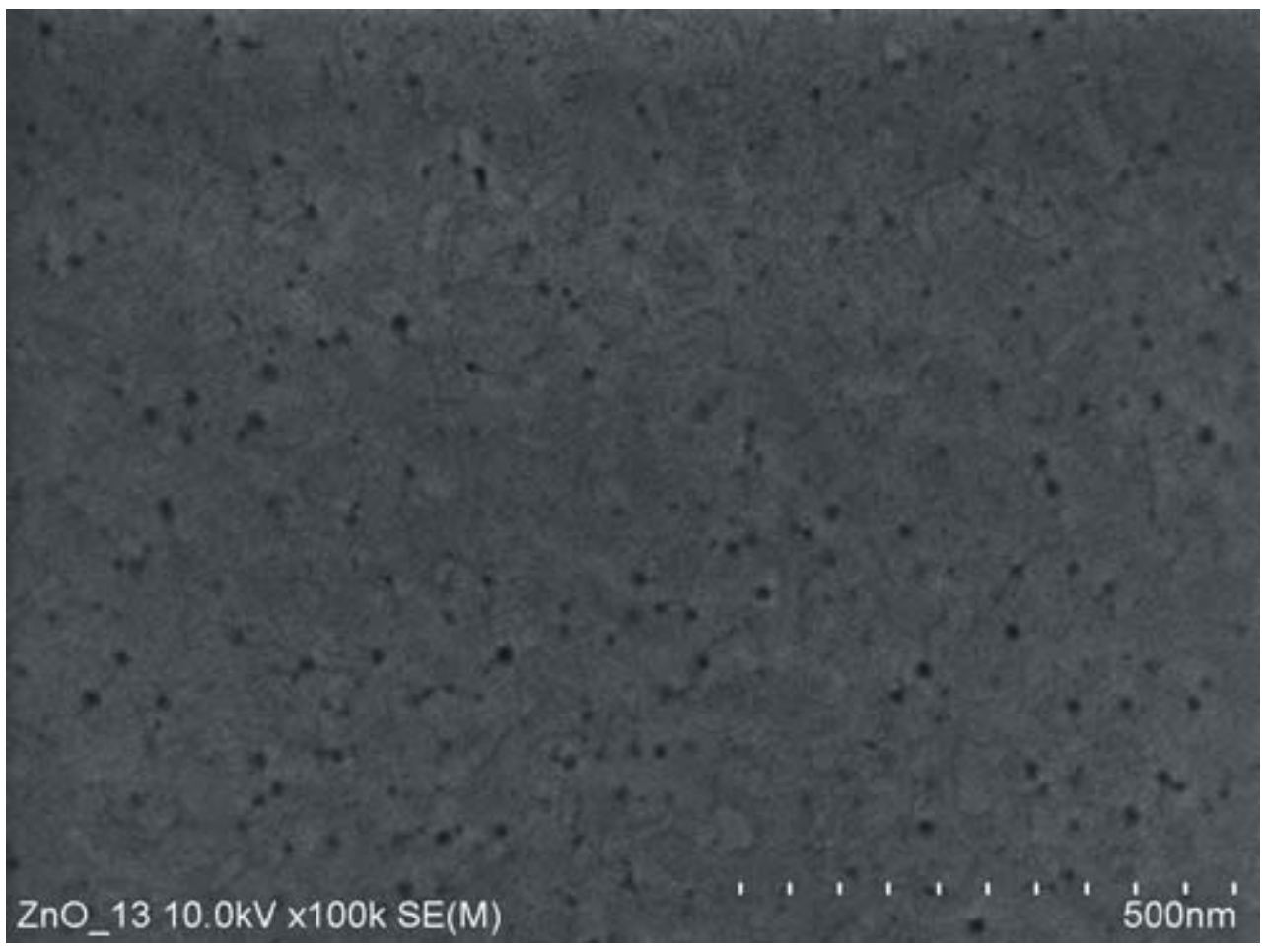
ZnO_13 10.0kV x100k SE(M)
500nm

(a)

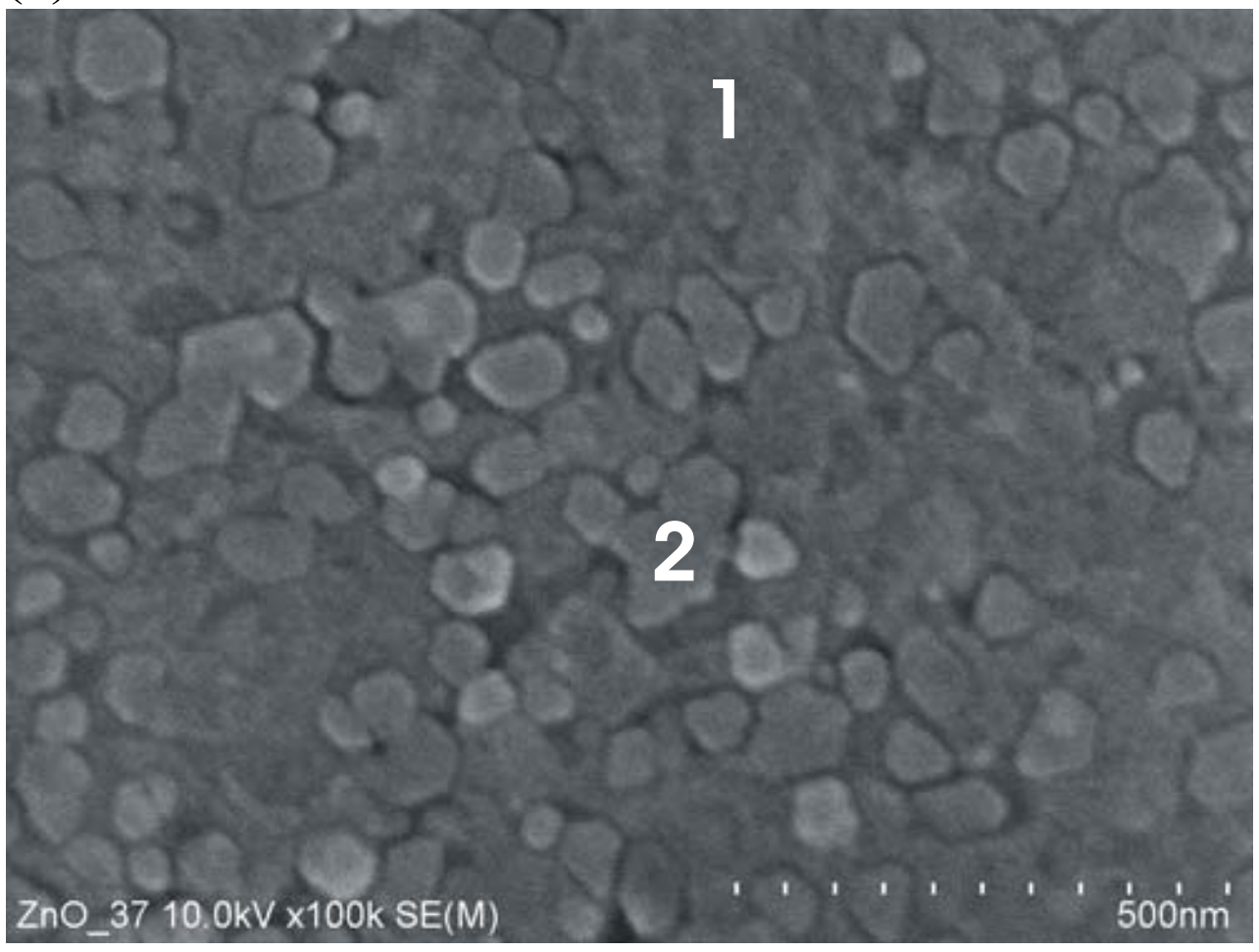
1
2
ZnO_37 10.0kV x100k SE(M)
500nm

(b)

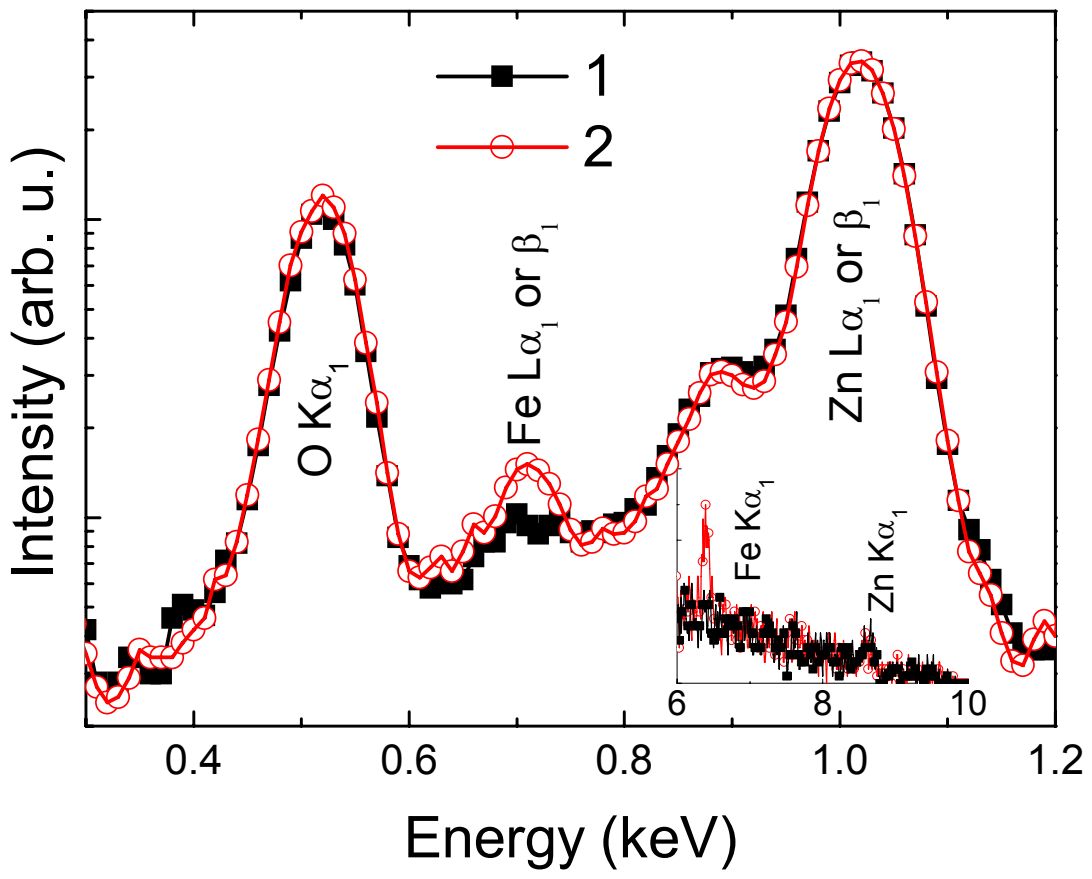

(c)

Fig. 7